\documentclass[11pt]{article}
\newcommand{\beq}{\begin{eqnarray}}
\newcommand{\eeq}{\end{eqnarray}}

\def\ltap{\ \raise.3ex\hbox{$<$\kern-.75em\lower1ex\hbox{$\sim$}}\ }
\def\gtap{\ \raise.3ex\hbox{$>$\kern-.75em\lower1ex\hbox{$\sim$}}\ }

\def\be{\begin{equation}}
\def\ee{\end{equation}}
\newcommand{\bel}[1]{\be\label{#1}}

\newcommand{\nomu}{{$\not\!\!\mu$SSM}}
\usepackage{latexsym}
\usepackage{epsfig}
\newcommand{\bc}{\begin{center}}
\newcommand{\ec}{\end{center}}
\newcommand{\y}{\nonumber \\}

\begin{document}
\begin{center}
{\large \bf The MSSM without $\mu$ term} 

\vspace{1cm}

{\bf 
Ann E. Nelson$^a$, Nuria Rius$^b$,  Veronica Sanz$^b$, Mithat Unsal$^a$}

$^a$Department of Physics, Box 1560, University of Washington, 
                 Seattle, WA 98195-1560, USA\\
$^b$Departamento de F\'\i sica Te\'orica and IFIC, Universidad de Valencia-
CSIC, \\
Edificio Institutos de Paterna, Apt. 22085, 46071 Valencia, Spain 

\vspace{1cm}

\begin{abstract}

We propose a supersymmetric extension of the standard model 
which does not have  a ``$\mu$''  supersymmetric Higgs mass parameter. 
The matter content of the MSSM is extended
with three additional chiral superfields: one singlet, 
an $SU(2)$ triplet and a color octet, and 
an approximate $U(1)_R$ symmetry naturally 
guarantees that $\tan\beta$ is large, 
explaining the top/bottom quark mass hierarchy.
Unlike in the MSSM, there are significant {\it upper} bounds on the
masses of superpartners, including an upper bound of 114 GeV on the
mass of the lightest chargino. However the MSSM bound on the lightest
Higgs mass does not apply.

\end{abstract}

\end{center}

\section{The \nomu\ and its low energy spectrum}
\label{spectrum}

In the Minimal Supersymmetric Model (MSSM) there is  a 
{\it supersymmetric} Higgs mass parameter,  ``$\mu$'',  which must be of order
of the electroweak scale for successful phenomenology. The difficulty 
of generating the correct mass scale for this 
supersymmetric mass parameter is the so called 
``$\mu$ problem''.   
This problem is more severe
in gauge mediated supersymmetry breaking (GMSB) models,  since 
it is quite difficult  in gauge mediation to induce a $\mu$ parameter which 
is naturally related to supersymmetry breaking,  without inducing an 
excessively large
$B\mu$ parameter \cite{Dine:1995vc}.

We consider an alternative solution to the $\mu$ problem, 
by building a viable model which does not have a $\mu$ parameter. 
In order to obtain a spectrum of superpartner masses  
experimentally acceptable without $\mu$ we have to add 
some matter content to the MSSM. 
This model, which we call the ``$\mu$-less 
Supersymmetric Standard Model'' (\nomu),  
has an approximate $U(1)_R$ symmetry which
guarantees naturally large $\tan\beta$, explaining the top/bottom 
quark mass hierarchy, and suppresses dangerous supersymmetric contributions to 
anomalous magnetic moments, $b\rightarrow s \gamma$, and proton decay.

The \nomu\ can naturally arise from either gauge or gravity 
mediation \cite{us}, 
if the supersymmetry breaking sector respects an approximate $U(1)_R$
symmetry. Such an approximate symmetry can easily arise by accident, as a 
consequence of the absence of  gauge singlet chiral superfields with 
$F-$terms in the supersymmetry breaking or mediation sector.

We  start with the principle that all mass terms arise directly either from 
electroweak symmetry breaking or from supersymmetry breaking. We therefore do 
not allow a supersymmetric $\mu$ term or any supersymmetric mass term.
The MSSM without a $\mu$ term  would have charginos lighter than the $W$ 
boson, which should have been found at LEP II, so we have to extend the 
theory.

In the exact $U(1)_R$ symmetric limit there are no supersymmetry breaking 
Majorana gaugino masses, so in order to give the gauginos Dirac masses 
we add three chiral superfields, namely a color octet $O$, a 
triplet under the $SU(2)$ gauge group, $T$ and a singlet $S$. 
These adjoint matter multiplets could have an extra dimensional origin, 
since extra dimensional theories in which gauge bosons live in the bulk and 
chiral matter fields live on a three brane typically have additional matter 
fields in the adjoint representation when described four dimensionally, 
unless the extra dimension is orbifolded. The adjoint fields might be 
$N=2$ superpartners of the gauge fields \cite{Fox}. 

\begin{figure} [t!] 
\label{pchar}
\begin{center}
\epsfig{figure=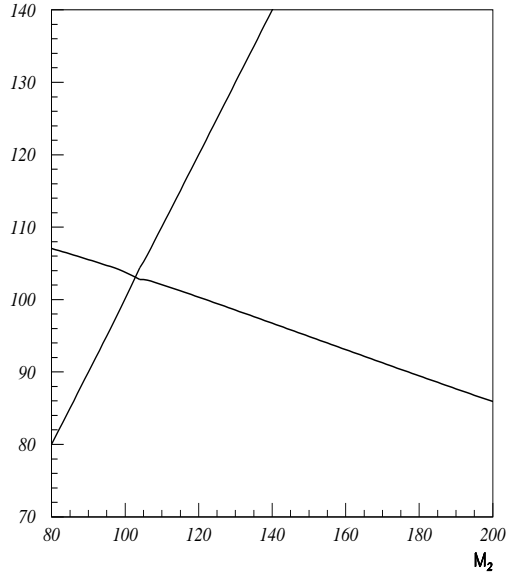,width=.5\textwidth,height=.4\textheight}
\end{center} 
\caption{Lighter chargino masses for $h_T=1$, $\tan \beta =60$ and 
$\tilde m_2=$5 GeV. }
\end{figure}

We now turn to a discussion of the spectrum of the \nomu, from the bottom up. 
The charge assignments of some of  the  components of Higgs and electroweak
gauge fields under the unbroken $U(1)_R$ are:
\begin{equation}
\begin{tabular}{|c|c|c|c|c|c|}
\hline
$\Psi_{H_1}$&
$\Psi_{H_2}$& 
$\Psi_T^{\pm}$&
${H_1}$&
${H_2}$&
$\lambda^{\pm}$\\
\hline
1&-1&-1&2&0&1  \\
\hline
\end{tabular}
\label{rcharge}
\end{equation}
Thus we can add the superpotential coupling 
\bel{tcoupling}
\int d^2\theta\ \  h_S S H_1 H_2  + h_T H_1 T H_2 \ . 
\ee
$U(1)_R$ charges are also assigned to quarks and leptons to allow the 
usual MSSM superpotential couplings.

Scalar trilinears involving the $T$ scalar are potentially troublesome, 
because they could induce a tadpole for $T$, which would get a vev
and lead to a large electroweak $T$ parameter. 
Sufficient suppression of this tree level contribution is provided 
by the approximate $U(1)_R$ symmetry and by a heavy mass 
for the $T$ scalar, which is automatic in the gauge mediated models.

Now the chargino mass matrix is 

\begin{equation}
\begin{tabular}{|r|c c c|}
\hline
  & $\Psi_T^+ $ & $-i \lambda^+  $ & $\Psi_{H_2}^+ $ \\
\hline
$\Psi_T^-  $ & $ 0$ & $\tilde M_2 $ & $ - 
h_T\, v_1$\\
$-i \lambda^-  $ & $ \tilde M_2$&$\tilde m_2$ & $\sqrt{2}\, m_W \,s_{\beta}$\\ 
$\Psi_{H_1}^-   $ & $h_T\, v_2$ &$\sqrt{2}\, m_W \,c_{\beta}  $&0\\ 
\hline
\end{tabular}
\label{chargino}
\end{equation}
where  $\tilde M_2$ ($\tilde m_2$) is a  soft supersymmetry breaking  
Dirac (Majorana) mass term. 
Note that  all the charginos can be made heavier than 104 GeV without a 
$\mu$ parameter.

With the $U(1)_R$ symmetry unbroken, $\tilde m_2 = v_1 = c_\beta =0$.
This will get modified slightly by small $U(1)_R$ breaking effects, which 
will get us away from the limit $\tan\beta\rightarrow\infty$ and set 
$\tan\beta$ to a moderate value $\sim 60$.
In this limit there is one chargino with mass $\tilde M_2$ and another chargino whose mass decreases with $\tilde M_2$. To obtain masses for all charginos 
heavier than 104 GeV,  while assuming $h_T < 1.2$,  $\tilde M_2$ must be in the range 104-120 GeV.
Moreover, the requirement, that all charginos should be heavier than 104 GeV
 leads to a lower bound on the Yukawa coupling, $h_T \gtap 1$.
Note that $\sqrt2 m_W=114$ GeV is an upper bound on the mass of the lightest 
chargino. 
Thus in the region where all charginos are heavier than 104 GeV we have two 
charginos with mass between 104 and 120 GeV and one heavier one. 
We show in Fig. 1 the lighter chargino masses as a function of 
$\tilde M_2$.

The neutralino mass matrix is:
\begin{equation}
\label{neutral}
\begin{tabular}{|r|c c c c c c|}
\hline
  & $\Psi_T^3$& $\Psi_S$&$-i \lambda'$ & $-i\lambda^3$ & $\Psi_{H_1}^1 $ &$\Psi_{H_2}^2$ \\
\hline
$\Psi_T^3$ & $0$ &0& $0$ & $ \tilde{M_2} $ & $h_T\, v_2/\sqrt{2}$ & 
$h_T \,v_1/\sqrt{2}$\\
$\Psi_S$&0&$0$&$\tilde{M_1}$&0&$h_S\,v_2/\sqrt{2}$&$h_S\,v_1/\sqrt{2}$\\
$-i \lambda'$ & $0$ &$\tilde{M_1}$& $\tilde m_1$ & $0$ & $-m_Z \, s_W \, c_{\beta}$ & $m_Z \, s_W \, s_{\beta} $\\
$-i\lambda^3$  &$ \tilde{M_2} $ &0& $0$ & $\tilde m_2$ & $m_Z \, c_W \, c_{\beta}$ & $-m_Z \, c_W \, s_{\beta} $\\
$\Psi_{H_1}^1$ & $h_T \, v_2/\sqrt{2}$ &$h_S\,v_2/\sqrt{2}$& $-m_Z \, s_W \, c_{\beta}$ & $m_Z \, c_W \, c_{\beta}$ & $0$ &
$0$\\ 
$\Psi_{H_2}^2$ & $h_T \, v_1/\sqrt{2}$ &$h_S\,v_1/\sqrt{2}$& $m_Z \, s_W \, s_{\beta}$ & $-m_Z \, c_W \, s_{\beta}$ & $0$ &
$0$\\ 
\hline
\end{tabular}
\end{equation}

\begin{figure} [t!] 
\label{pneu}
\begin{center}
\epsfig{figure=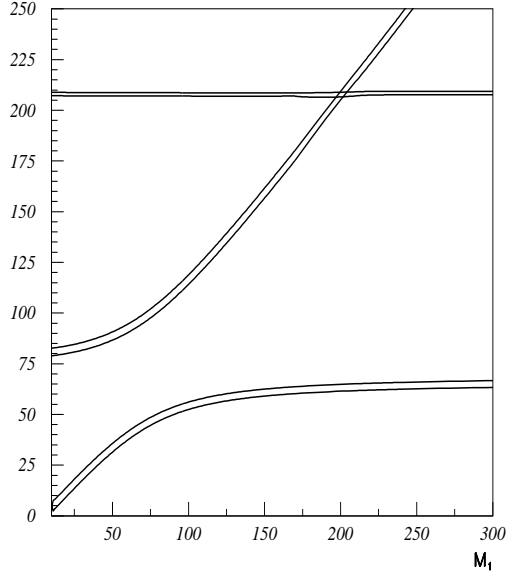,width=.5\textwidth,height=.4\textheight}
\end{center} 
\caption{Neutralino masses as a function of $\tilde M_1$, for 
$\tilde M_2 = 104$ GeV, $h_T = 1$, $h_S = 0.1$, $\tan \beta=60$ and 
$\tilde m_1=\tilde m_2=$5 GeV.}
\end{figure}

In the large $\tan\beta$, $U(1)_R$ symmetric limit the masses become 
approximately Dirac.
There is always a nearly Dirac neutralino with mass 
lighter than the $Z$.
In Fig. 2 we show the  neutralino masses as a function of the
soft mass term $\tilde M_1$, for $h_T = 1$ and $h_S = 0.1$.
In principle the Yukawa coupling $h_S$ is a free parameter, 
but large values are disfavored by electroweak precision measurements.

Similarly, gluinos get a supersymmetry breaking  Dirac mass term
by mixing with the fermionic component of the color octet $O$.
The scalar superpartners receive soft supersymmetry breaking masses as
usual. 
A very small scalar $\mu B$ term of order a few GeV$^2$
\be \mu B H_1 H_2\ee will be needed in order to induce a small vev for
$H_1$, which gives the leptons and down-type quarks mass.
It is natural for this term to be small as it breaks the
approximate $U(1)_R$ symmetry. Because the symmetry is explicitly broken 
rather than spontaneously broken, there is no light pseudoscalar. 

The MSSM bound on the lightest Higgs mass does not apply, though, since 
the scalar sector is also enlarged by the scalar components of the 
$SU(2)$ triplet and scalar chiral superfields, and there are new, 
$F-$component contributions to the Higgs quartic coupling. 
There will still be some upper bounds, computed
in  general models with Higgs triplets in refs. 
\cite{Espinosa}.


\section{$U(1)_R$ Symmetric Gauge Mediation}
\label{rsgmsb}

We assume that supersymmetry breaking is transmitted to
the \nomu \ by Gauge Mediated Supersymmetry Breaking (GMSB), and 
a messenger sector of heavy supermultiplets in a vector-like
representation of the standard gauge group. 
In conventional gauge mediation, the messengers learn about
supersymmetry breaking from coupling to a gauge singlet with an $F-$term.
This transmits both supersymmetry breaking and $U(1)_R$ symmetry
breaking to the MSSM. Since we want an approximately $U(1)_R$
symmetric \nomu, we will assume the messenger sector does not contain any 
singlet.  Instead supersymmetry breaking in the
messenger sector is primarily mediated by some new
gauge group also carried by the messengers. Such mediation will primarily 
induce nonholomorphic scalar supersymmetry breaking masses in the messenger 
sector \cite{Randall:1997zi,Csaki:1998if}. 

We  assume the usual 
messenger matter content of chiral superfields $L,\bar L, D,\bar D$
where $L,\bar L$ transform under $SU(2)\otimes U(1)$ in conjugate
representations  and $D, \bar D$
carry color.  
In order to obtain Dirac gaugino masses, $S$, $T$ and $O$ must couple to
the messengers. The messenger superpotential   
is
\bel{mess}
\lambda_S S \bar L L+\lambda'_S S \bar D D +
\lambda_T T \bar L L+ \lambda_O O\bar D D + M_L \bar L L + M_D \bar D D
\ .
\ee
The supersymmetric mass parameters $M_L$ and $M_D$, which can be 
dynamically generated \cite{Csaki:1998if}, 
are much heavier than the weak scale. 

The mass matrix for, {\it e.g.} the $L, \bar L$ scalar fields will
have the following  form 
\bel{messass}
\pmatrix{
M_L^2 + \tilde m_L^2 & 0\cr  
0 & M_L^2 +\tilde m_{\bar L}^2  
}
\ee
where $\tilde m_L^2,\tilde m_{\bar L}^2$ are soft supersymmetry
breaking masses. 
With no messenger singlet, to leading order the messenger sector 
will accidentally have unbroken $U(1)_R$ symmetry, and no
Majorana gaugino masses will be produced. However, at one loop, 
the gauginos couple to the fermionic components of $T,O$ and $S$ and get a 
Dirac supersymmetry breaking mass.

 Note also that provided the $D-$type masses are generated by new gauge interactions whose generators are orthogonal to electroweak hypercharge,  {\it i.e.}
${\rm Tr}\ T_Y T_{\rm new}=0\ ,$ 
the  disaster of generating a $D$-term for hypercharge at one 
loop is avoided.

There are two diagrams contributing to Dirac gaugino masses, which cancel in
the limit that  $\tilde M_{L,D}^2= \tilde M_{\bar L,\bar D}^2$.
In the limit  that the supersymmetry breaking terms are much smaller
than $M_L$, the Dirac masses $\tilde M_{2,3}$ are
\be
\tilde M_{2,3}=S_{L,D}   \frac{g_{2,3} \lambda_{T,O}}{4\pi^2} { \tilde m_{\bar
    L, \bar D}^2-\tilde m_{L,D}^2\over M_{L,D}}\ .
\ee
where $S_{L,D}$ are the Dynkin indices of the
$L,D$ representations respectively.  Similarly, $\tilde M_1$ will receive contributions from both $L$ and $D$.

The masses of scalar \nomu\ particles may be found as a special case of
the general expressions computed in \cite{Poppitz}.
Note that obtaining positive squark and slepton  masses will require negative
supertrace in the messenger sector. 
As a consequence,  
the scalar components of $T,S,$ and $O$ will receive a large positive mass 
squared at one loop and will therefore be significantly heavier than the 
other superpartners. This mass is of order a loop factor times the soft masses
in the messenger sector, and is not suppressed by the messenger mass scale. 
The $T$ and $O$ scalar masses  should not be much larger than
$10^4$ GeV, or they  will give excessive two loop  contributions to squark 
and slepton masses.  The supersymmetry breaking terms in the messenger sector 
should therefore not be larger than of order $M_S\sim 10^5$ GeV.  Since 
squark and slepton masses will be of order $(\alpha/\pi)(M_S^2/M)$, the 
messenger mass scale $M$ should  be below $10^6$ GeV.

The \nomu\ avoids the gauge mediated $\mu$ problem, because 
a  $\mu B$ parameter can be induced which is proportional to a small coupling,
and it is not a problem that the resulting $\mu$ parameter will be much 
smaller than the weak scale.

\section{Contribution to precision electroweak parameters}
\subsection{$T$ parameter}
\label{T}

The 
superpotential couplings $h_T T H_1 H_2$ and $h_S S H_1 H_2$
break custodial $SU(2)$ symmetry and thus can lead to potentially large 
one-loop effects in the $T$ parameter. 
Although the oblique approximation is not appropriate for light 
superpartners, we
shall interpret our results for the $T$ parameter as an order of
magnitude estimate of the radiative corrections expected in the 
\nomu.

We find that the leading contribution to the $T$ 
parameter grows as $h_T^2 \log (h_T^2 v^2/\mu^2)$ and it is 
therefore very sensitive to the exact value of the coupling $h_T$.
Recall that there is a lower limit on this coupling from chargino
masses. Although the singlet coupling $h_S$ also contributes
to the $T$ parameter, its contribution is negligible provided 
$h_S \ltap 0.1$ \footnote{There is also  
a $T$ parameter contribution from the scalar sector, 
but it is not enhanced by the $\log \mu^2$ term,
and can be made very small by the soft supersymmetry breaking scalar
masses.}.

From a global fit of the electroweak precision data one obtains 
$T=-0.02 \pm 0.13 (+0.09)$, where the central value assumes 
$M_H= 115$ GeV and the parentheses shows the change for $M_H =300$
GeV \cite{Groom:2000in}. 
This bound can be relaxed for larger $M_H$, leading to
$T \ltap 0.6$ at 95\% CL \cite{Chivukula}. 
If we impose the kinematic limit from LEP II that charginos should be
heavier than 104 GeV, $h_T \sim 1$ and the contribution to the $T$ 
parameter is  huge, $\sim$ 2.7.
However if the actual bound on chargino masses in this model were 
somewhat lower, say 90 GeV, we would obtain $h_T \gtap 0.6$
which leads to $T \sim 0.6$. 
Therefore, given the large sensitivity of the $T$ parameter to the 
value of $h_T$, a careful calculation of the chargino mass bounds  
is crucial to determine the viability of the \nomu\ model.

\subsection{Muon anomalous magnetic moment}
\label{suppress}

The supersymmetric contributions to $a_\mu$ 
\cite{Moroi} include  
loops with a chargino and a muon sneutrino and loops with a 
neutralino and a smuon. 
Explicit formulae can be found in \cite{us}.
In the $U(1)_R$ symmetric limit, the contributions to $a_\mu$ proportional 
to the neutralino and chargino masses exactly vanish, and there is only
a tiny effect proportional to $m_\mu$. However, 
once we take into account the small $U(1)_R$ symmetry breaking effects,
the leading contribution comes from the terms with the neutralino and 
chargino masses, much as in the MSSM.
The contribution from chargino loops is typically dominant, 
except for $m_L \gg m_R$.

\begin{figure} [ht!] 
\label{gminustwo}
\begin{center}
\epsfig{figure=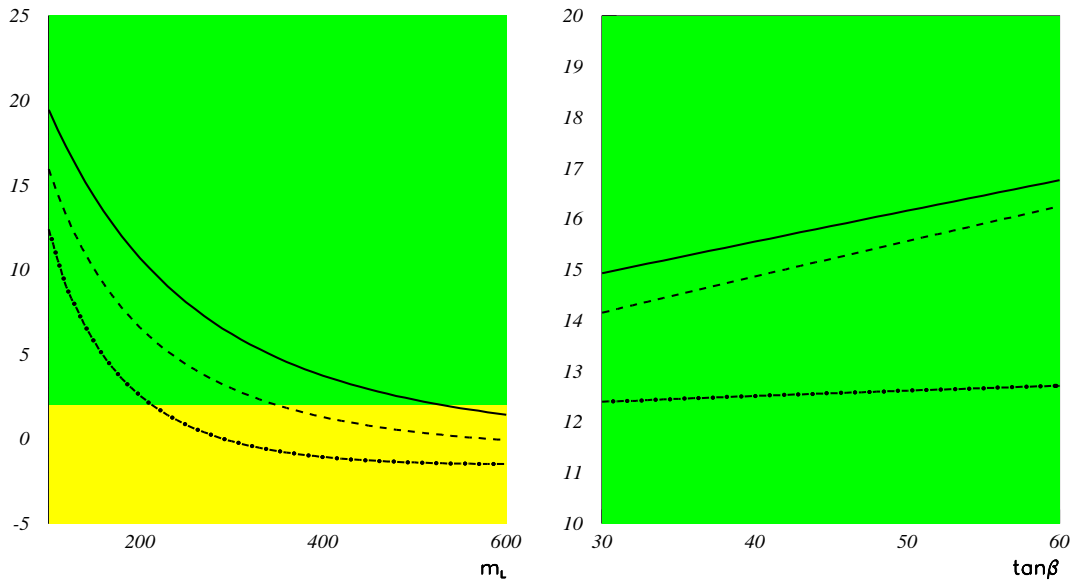,width=\textwidth,height=.4\textheight}
\end{center} 
\caption{Maximum value of $\delta a_{\mu} \times 10^{10}$ as a function of 
$m_L$ and $\tan \beta$,  
for
$ \tilde{m}_1 =\tilde{m}_2$ = 0 (dashed-dotted), 
5 GeV (dashed) and 10 GeV (solid).
We have taken $A=0$, $m_R =$ 100 GeV, $\tilde{M}_1=100$ GeV,  
$\tilde{M}_2=110$ GeV, $h_T = 0.8,\ h_S=0.1$, 
 $\tan\beta=60$ (left) and $m_L =$ 100 GeV (right).
The shadowed areas correspond to 1$\sigma$ (dark-green) and 2$\sigma$ 
(light-yellow) allowed regions from the $g-2$ collaboration result.}
\end{figure}

In Fig. 3 we show the maximum possible value of 
$\delta a_\mu$ in the \nomu\ model as a function of the soft 
supersymmetry breaking mass term $m_L$ (left) and 
as a function of $\tan \beta$ (right),  
for several values of the gaugino Majorana masses. 
Although the contribution to $a_\mu$ in the \nomu\ model 
is also enhanced for large $\tan\beta$, due to the approximate $U(1)_R$ it 
is suppressed by the small
gaugino Majorana masses and therefore much smaller than in the
MSSM.

\section{Unification of couplings}
\label{unify}
One rational for
supersymmetry is coupling constant unification. If we add matter to
the MSSM in incomplete multiplets under the unifying group
the usual successful prediction of $s_W^2\approx .23$ may be lost. 
In the \nomu\ we have added matter in the adjoint representation of
$U(1)\otimes SU(2)\otimes SU(3)$, which will not preserve the usual  
prediction. It is, however a simple matter to embed the $T,S$ and $O$ fields 
into a complete
adjoint multiplet of a GUT such as $SU(3)^3$ or $SU(5)$.

Although it is not necessary, the other fields of the multiplet can serve as 
the messenger fields of a gauge mediated model \cite{us}. 
If one assumes all the \nomu\ superpartners are at the weak scale,
and computes the one-loop running neglecting threshold effects, one
can  fit the scales of the new matter multiplet and GUT  to the low energy
gauge coupling constants. The result is
\beq
M_{\rm new}&=&M_{\rm weak}e^{{2\pi\over
    3}\left({12\over\alpha_2}-{5\over\alpha_1}-{7\over\alpha_3}\right)}\y \\
M_{\rm GUT}&=&M_{\rm weak}e^{{5\pi\over
    6}\left({3\over\alpha_2}-{1\over\alpha_1}-{2\over\alpha_3}\right)}\ .
\eeq
By taking values for the coupling constants at the edge of their allowed
ranges, {\it e.g.} $\alpha(M_Z)$ = 1/127.7, $\alpha_s(M_Z)$ =
0.122, and $s_W^2$ = 0.233 the additional matter fields can be as heavy as 
$3\times 10^7$ GeV
and the GUT scale as high as $10^{18}$ GeV. Threshold effects at the GUT, 
messenger
and \nomu\ scales  and higher loop corrections   make order one
changes in these predictions. This constraint is less stringent than the 
upper bound on the messenger scale found in section~\ref{rsgmsb}.

\vskip 0.25in
{\bf Acknowledgments} 
\vskip 0.15in
This work was partially 
supported by the DOE under contract DE-FGO3-96-ER40956, by the  
Spanish MCyT grants PB98-0693 and FPA2001-3031, 
by ERDF funds from the European Commission, and by  
the TMR network contract HPRN-CT-2000-00148 of the European Union.

\end{document}